\documentclass[sigconf, 9pt]{acmart}

\AtBeginDocument{%
  \providecommand\BibTeX{{%
    \normalfont B\kern-0.5em{\scshape i\kern-0.25em b}\kern-0.8em\TeX}}}

\copyrightyear{2024} 
\acmYear{2024} 
\setcopyright{rightsretained} 
\acmConference[ICCAD '24]{IEEE/ACM International Conference on Computer-Aided Design}{October 27--31, 2024}{New York, NY, USA}
\acmBooktitle{IEEE/ACM International Conference on Computer-Aided Design (ICCAD '24), October 27--31, 2024, New York, NY, USA}
\acmDOI{10.1145/3676536.3676777}
\acmISBN{979-8-4007-1077-3/24/10}


\usepackage{multirow}
\usepackage{algorithm}
\usepackage{graphicx}
\usepackage{textcomp}
\usepackage{xcolor}
\usepackage{tabularx}
\usepackage{stfloats}
\usepackage{lipsum}
\usepackage{pgfplots}
\usepackage{pgf-pie}
\usepackage{algpseudocode}
\usepackage{pgfplotstable}
\usepackage{subfig}
\usepackage[colorinlistoftodos]{todonotes}
\usepackage{enumitem}
\pgfplotsset{compat=1.17}
\newcommand{\ignore}[1]{}

\newif\ifremark
\long\def\remark#1{
\ifremark%
        \begingroup%
        \dimen0=\columnwidth
        \advance\dimen0 by -1in%
        \setbox0=\hbox{\parbox[b]{\dimen0}{\protect\em #1}}
        \dimen1=\ht0\advance\dimen1 by 2pt%
        \dimen2=\dp0\advance\dimen2 by 2pt%
        \vskip 0.25pt%
        \hbox to \columnwidth{%
                \vrule height\dimen1 width 3pt depth\dimen2%
                \hss\copy0\hss%
                \vrule height\dimen1 width 3pt depth\dimen2%
        }%
        \endgroup%
\fi}

\remarktrue

\begin{document}

\title{Sustainable Hardware Specialization}
\author{*Pranav Dangi$^\dag$, *Thilini Kaushalya Bandara$^\dag$, Saeideh Sheikhpour$^\ddagger$}
\author{Tulika Mitra$^\dag$ and Lieven Eeckhout$^\ddagger$\vspace{1mm}}
\thanks{*Equal Contribution}
\affiliation{%
  \institution{$^\dag$National University of Singapore and $^\ddagger$Ghent University, Belgium}
  \country{}
}


\begin{abstract}
Hardware specialization is commonly viewed as a way to scale performance in the dark silicon era with modern-day SoCs featuring multiple tens of dedicated accelerators. By only powering on hardware circuitry when needed, accelerators fundamentally trade off chip area for power efficiency. Dark silicon however comes with a severe downside, namely its environmental footprint. While hardware specialization typically reduces the operational footprint through high energy efficiency, the embodied footprint incurred by integrating additional accelerators on chip leads to a net overall increase in environmental footprint, which has led prior work to conclude that dark silicon is not a sustainable design paradigm.

We explore sustainable hardware specialization through reconfigurable logic that has the potential to drastically reduce the environmental footprint compared to a sea of accelerators by amortizing its embodied footprint across multiple applications. We present an abstract analytical model that evaluates the sustainability implications of replacing dedicated accelerators with a reconfigurable accelerator. We derive hardware synthesis results on ASIC and CGRA (a representative reconfigurable fabric) for chip area and energy numbers for a wide variety of kernels. We input these results to the analytical model and conclude that reconfigurable fabric is more sustainable. 
We find that as few as a handful to a dozen accelerators can be replaced by a CGRA. Moreover, replacing a sea of accelerators with a CGRA leads to a drastically reduced environmental footprint (by a factor of $2.5 \times$ to $7.6 \times$).
\end{abstract}

\thispagestyle{empty}
\pagestyle{empty}

\maketitle

\section{Introduction}
The end of Dennard scaling~\cite{dennard} has dramatically changed how we design processors. Increased power density as we transition to new chip technology nodes leads to dark silicon~\cite{dark-silicon,dark-silicon-hadi}, which means that we cannot power on the entire chip while keeping thermals within a safe operating range. Hardware specialization enables continuous performance scaling despite dark silicon by executing specific kernels on dedicated hardware accelerators. Powering on an accelerator provides high performance when needed; when not in use, an accelerator is powered off to save power. The advent of dark silicon has created a flurry of work in hardware specialization, sometimes referred to as the `golden age for computer architecture'~\cite{golden-age}, with accelerators for machine learning, video coding and decoding, image signal processing, security encryption/decryption, etc. In fact, a modern-day computer is a system-on-chip (SoC) in which general-purpose CPU and GPU cores are complemented with a sea of (multiple tens of) domain-specific accelerators (DSAs)~\cite{hill-reddi}. This is the case across the computing spectrum from mobile application processors (e.g., Qualcomm Snapdragon~\cite{snapdragon}) to laptop and desktop processors (e.g., Apple M2~\cite{apple-M2}, Intel Sapphire Rapids~\cite{intel-sapphire-rapids}) and server processors (e.g., AMD EPYC~\cite{amd-epyc}, IBM Telum~\cite{ibm-telum}). The number of DSAs integrated on chip has steadily increased over time: Shao et al.~\cite{aladdin-retrospective} report that for Apple SoCs, the number of DSAs has increased from less than 10 in the A4 (2010) to more than 40 in the A12 (2018).

Fundamentally, dark silicon trades off chip area for power efficiency, i.e., 
hardware specialization boosts performance within the available power budget by powering on hardware resources only when needed. Dark silicon thus comes at the cost of additional transistors to implement the various DSAs on chip. Fortunately, transistors have become exponentially cheaper over time thanks to Moore’s Law~\cite{bohr}, which makes dark silicon economically viable (even today, despite Moore’s Law slowing down). In other words, hardware specialization offers continuous performance scaling without increasing the area (and cost) per chip.

\begin{figure}[tb]
\centering
\includegraphics[width=0.9\columnwidth]{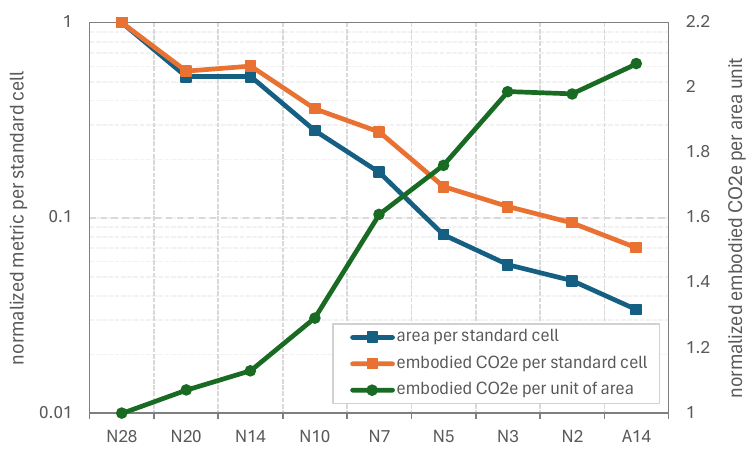}
\vspace{-2mm}
\caption{Chip area and embodied footprint per standard cell (left axis) and embodied footprint per unit of chip area (right axis) for various chip technology nodes normalized to 28\,nm~\cite{imec,imec-2023}. {\it Keeping chip area constant to accommodate dark silicon comes at the cost of an increased embodied footprint.} }
\label{fig:standard-cell-footprint}
\end{figure}

There is a severe downside to dark silicon though, which has been largely ignored, namely its environmental footprint. With information and communication technology (ICT) being responsible for 2.1\% to 3.9\% of the global greenhouse gas (GHG) emissions world-wide~\cite{Freitag} — currently on par with the aviation industry, and it is projected to continue to grow~\cite{eeckhout2023}. Figure~\ref{fig:standard-cell-footprint} (vertical axis on the left) reports chip area and the embodied carbon footprint (i.e., environmental footprint due to manufacturing, measured in CO2e equivalent) per standard cell for nine available and projected technology nodes as provided by imec~\cite{imec,imec-2023}, normalized to 28\,nm. Continuous advancements in chip technology have dramatically reduced chip area as well as the embodied footprint per standard cell. However, the embodied footprint does not reduce at a similar pace as chip area due to increased complexity in manufacturing, i.e., more processing steps leading to increased energy consumption and GHG emissions. This implies that for a constant unit of chip area, the embodied footprint for logic has substantially increased, see Figure~\ref{fig:standard-cell-footprint} (vertical axis on the right). In other words, dark silicon comes at the cost of a substantially increased embodied footprint.

The question now is whether the increase in embodied footprint is offset by the decrease in operational footprint due to device use during its entire lifetime. While hardware specialization typically reduces energy consumption (due to higher performance and/or lower power) when in use, thereby reducing the operational footprint, the embodied footprint for integrating the DSA on chip has to be incurred regardless. This suggests that dark silicon only leads to a net reduction in environmental footprint if the DSAs are frequently used, which seems to contradict the notion of dark silicon. This has led prior work to suggest that dark silicon is not sustainable from an environmental perspective~\cite{dark-silicon-harmful,focal}. 
\begin{figure}[tb]
\centering
\includegraphics[width=0.9\columnwidth]{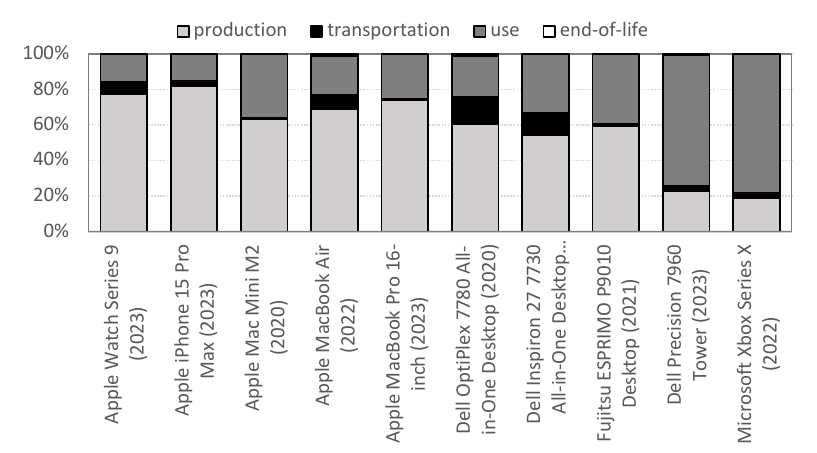}
\vspace{-4mm}
\caption{Breakdown of the environmental footprint in production, transportation, use and end-of-life processing. {\it The embodied footprint dominates for most computing devices.} }
\label{fig:emb-vs-op-ratio}
\vspace{0mm}
\end{figure}
In this work, we explore a sustainable alternative to dark silicon, namely hardware specialization through reconfigurable logic. The intuition that underpins this work is that a fabric that can be dynamically reconfigured or reprogrammed across applications can possibly achieve the best of both worlds, i.e., yield high performance at high power efficiency without incurring the vast embodied footprint of dark silicon. Replacing the sea of DSAs by a reconfigurable fabric with overall smaller area has the potential to drastically reduce the embodied footprint. On the flip side, a reconfigurable fabric is less efficient and leads to increased energy consumption and thus a higher operational footprint. The fundamental question hence is whether the decrease in embodied footprint offsets the increase in operational footprint --- if so, reconfigurable logic is a more sustainable design paradigm than dark silicon. 

To investigate the opportunity for sustainable hardware specialization through reconfigurable logic, we formulate an abstract analytical model to compare the environmental footprint of a sea of DSAs versus a reconfigurable fabric. The model determines the {\it critical DSA count (CDC)} or the number of DSAs the reconfigurable fabric needs to replace for the latter to be more sustainable. Despite the model being parameterized to account for inherent data uncertainty, several interesting insights can be obtained. First, we find that --- contrary to common belief  --- area efficiency is more important than energy efficiency for a DSA to be sustainable. Second, CDC decreases (1) with an increasing contribution of the embodied footprint in the overall environmental footprint of a device, (2) with a decreasing area (and to a lesser extent the energy) efficiency of a DSA relative to the reconfigurable fabric, and (3) with a decreasing degree of DSA concurrency during application execution.


To quantify the area and energy efficiency of a DSA versus a reconfigurable fabric, we map a total of eight widely used application kernels to DSA as well as reconfigurable fabric while assuming iso-performance implementations. We consider standard-cell ASIC implementations for the DSA, and coarse-grain reconfigurable array (CGRA) ~\cite{hycube, softbrain, plasticine, snafu} for the reconfigurable fabric. We report that a DSA incurs on average 0.27$\times$ and 0.31$\times$ less chip area and energy compared to CGRA, respectively. 
This suggests that for embodied footprint dominated systems, the carbon footprint of CGRA equals that of 4 to 5 DSAs. In a system comprising a total of 40 DSAs, this reduces the environmental footprint by a factor of 2.5$\times$ to 7.6$\times$.
The overall conclusion is that CGRA is a sweet spot, paving a way towards sustainable hardware specialization.

\vspace*{-0.2em}
\section{Analytical Carbon Modeling}
\vspace*{-1pt}
\subsection{Background}



The environmental footprint for a computing device consists of two major contributors~\cite{greenchip}: (1) the {\it embodied footprint} due to raw material extraction, manufacturing, assembly, transportation, end-of-life processing, and (2) the {\it operational footprint} due to device use during its entire lifetime. Figure~\ref{fig:emb-vs-op-ratio} reports the breakdown of the environmental footprint for various computing devices including a smart watch, smartphone, laptop, medium-end and high-end desktop computers as well as a gaming console from different vendors including Apple, Dell, Fujitsu and Microsoft obtained from their corresponding product environmental reports. 

The environmental footprint is dominated by the embodied footprint for many computing devices, especially mobile devices such as laptops (embodied footprint accounts for 70--75\% of total) and smartphones and smart watches (80--85\%). In contrast, high-end desktop computers and gaming consoles are mostly dominated by the operational footprint, i.e., the embodied footprint contributes for 20--25\%. Medium-end desktops are somewhere in the middle for which the embodied footprint ranges between 55--60\%. This is in line with a prior survey by Gupta et al.~\cite{gupta-hpca2021} which concluded that battery-operated consumer electronics such as smart watches, phones, tablets and laptops are mostly dominated by the embodied footprint, whereas always-connected devices such as desktops and gaming consoles are mostly dominated by the operational footprint. 


\subsection{Abstract Carbon Model}

Analyzing the environmental footprint of a chip is difficult due to inherent data uncertainty~\cite{focal,ACT} resulting from semiconductor manufacturing industry secrecy, unknown or poorly documented material supply chains, hard to anticipate product usage and lifetimes, etc. To make things even more complicated, rebound effects may turn per-device efficiency improvements into increased usage and deployment, ultimately increasing rather than decreasing the overall environmental footprint, referred to as Jevons' paradox~\cite{jevons}.

The degree of uncertainty calls for an abstract parameterized analytical model based on first-order principles. We rely on the same proxies used in the FOCAL model~\cite{focal} to quantify the embodied and operational footprint, and a parameter $\alpha_{E2O}$ ($0 \leq \alpha_{E2O} \leq 1$) to weigh the relative importance of the embodied and operational footprint. 
The proxy for the embodied footprint is chip area, while the proxy for the operational footprint is energy consumption.\footnote{This assumes a fixed-work scenario. The FOCAL model also includes a fixed-time scenario for which the proxy for the operational footprint is power consumption. This is not considered here as we assume iso-performance design alternatives in this work.} 
As reported previously, the embodied-to-operational ratio $\alpha_{E2O}$ varies across computing devices: 0.7--0.85 for battery-operated devices (watches, smartphones, laptops) versus 0.2--0.6 for always-on devices (desktops, gaming consoles).

We now construct an abstract analytical model to compare the environmental footprint of a sea of DSAs versus the reconfigurable fabric. We initially assume that only a single DSA is used at any given time; we relax this assumption later.

\noindent{\bf Serial DSA usage.}
We first assume that only one DSA is active at any time, i.e., DSAs are used serially. The reconfigurable fabric incurs a smaller environmental footprint than a sea of DSAs if\vspace*{-0.08in}
\begin{equation}
\alpha_{E2O} \cdot N \cdot A + (1 - \alpha_{E2O}) \cdot E > 1,
\end{equation}
with $N$ the number of DSAs, and $A$ and $E$ the (average) chip area and energy consumption per DSA relative to the reconfigurable fabric, respectively. In other words, if the weighted embodied footprint of integrating $N$ DSAs (first term) plus the weighted operational footprint of using a DSA (second term) is larger than one, this implies that the sea of DSAs incurs a larger environmental footprint than a reconfigurable fabric with a normalized environmental footprint equal to one. The reconfigurable fabric is assumed to be large enough to accommodate the largest kernel, i.e., it has sufficient resources available to execute any single kernel in its entirety. 

\begin{figure*}[tb]
\centering
\subfloat[\small serial DSA execution ($n=1$)]{
\includegraphics[width=0.8\columnwidth]{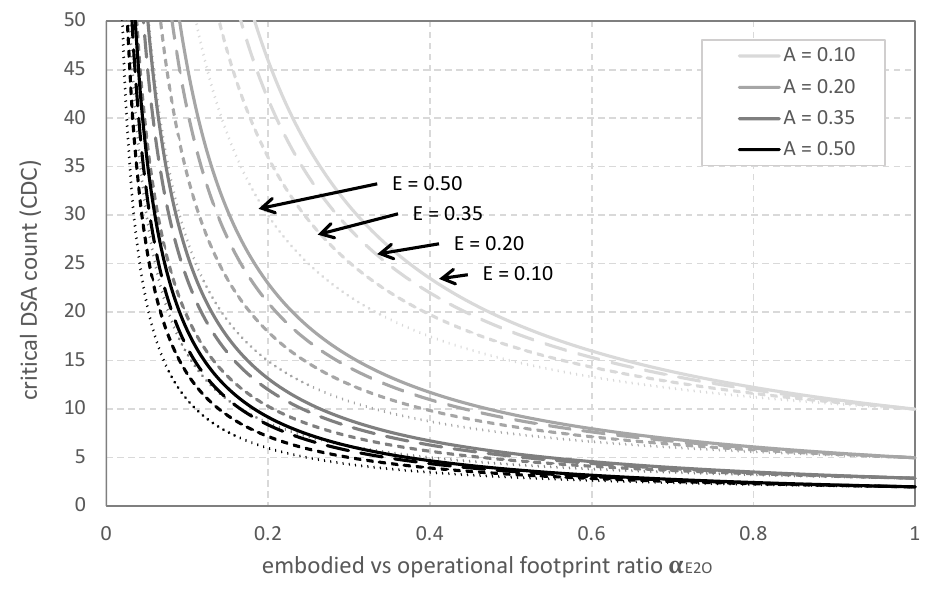}}\hspace{4mm}
\subfloat[\small concurrent DSA execution ($n=3$)]{
\includegraphics[width=0.8\columnwidth]{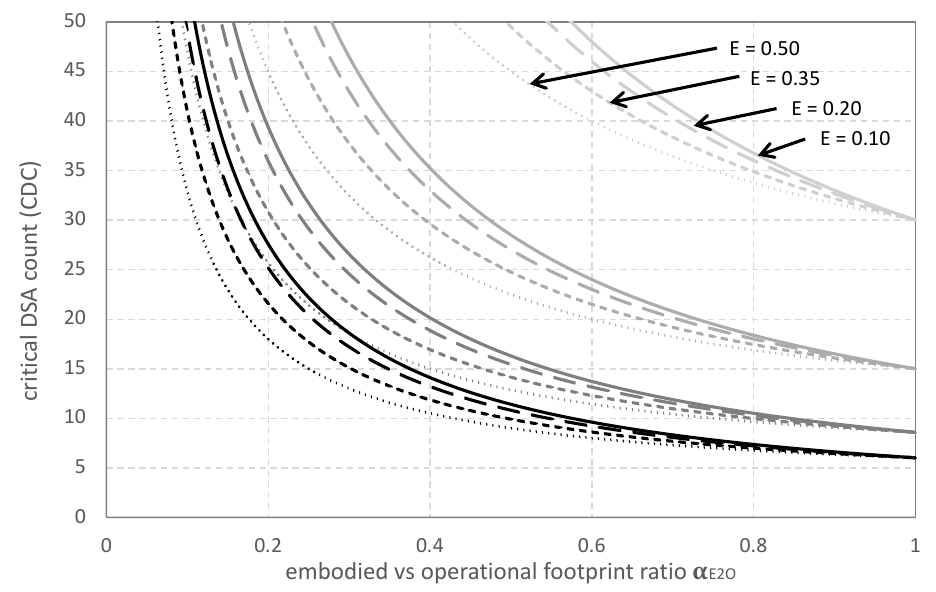}}
\vspace{-2mm}
\caption{Critical DSA count (CDC) as a function of $\alpha_{E2O}$, $A$ and $E$: assuming (a) serial DSA execution ($n=1$) and (b) concurrent DSA execution ($n=3$). {\it CDC decreases with increasing $\alpha_{E2O}$, $A$ and $E$, suggesting that a reconfigurable fabric has the potential to be more environmentally friendly than a sea of DSAs if the embodied footprint is substantial and if the area (and energy) efficiency gain of dedicated DSAs is limited relative to the reconfigurable fabric.}}
\label{fig:CDC}
\vspace{-3mm}
\end{figure*}

\vspace{1mm}
\noindent{\bf Multiple concurrent DSAs.}
While some applications exercise a single DSA at a time, others exercise multiple concurrently, for example, in a streaming fashion whereby one DSA produces input for the next DSA to process. Many modern-day applications exercise several different DSAs concurrently~\cite{hill-reddi}. The number of concurrently exercised DSAs is typically small though, at most a handful, see also Table~\ref{tbl:kernels} as reported by several prior works~\cite{hill-reddi,earable}, which --- fortunately --- aligns well with the restrictions imposed by dark silicon. We extend the serial DSA usage model to concurrent execution as follows. The reconfigurable fabric is more sustainable than a sea of DSAs if
\vspace*{-0.08in}
\begin{equation}
\alpha_{E2O} \cdot N \cdot A + (1 - \alpha_{E2O}) \cdot n \cdot E > n,
\end{equation}
with $n$ the number of DSAs exercised concurrently during runtime. This model considers that the operational footprint of the DSAs equals $n \cdot E$. The model further conservatively assumes that the footprint of the reconfigurable fabric is $n$ times larger if it needs to execute $n$ kernels concurrently. This is a conservative estimate because a reconfigurable fabric can inherently reuse hardware resources across kernels, such as sharing memory. Moreover, a reconfigurable fabric designed to accommodate the largest kernel is typically able to spatially accommodate multiple smaller kernels simultaneously, thereby reducing both the area overhead and the energy overhead compared to the linear scaling assumption. This inherent resource-sharing and spatial mapping flexibility of the reconfigurable fabric allows for more efficient execution of multiple kernels, leading to a lower overall area/energy overhead than the conservative estimate used in the model.

\begin{table}[t]
\centering
\caption{Number of (concurrent) kernels per application.}\label{tbl:kernels}
\vspace{-2mm}
\resizebox{0.9\columnwidth}{!}{
\small
\begin{tabular}{|l|c|l|}
\hline
{\bf Reference} & {\bf No.\ kernels} &{\bf Application domain} \\ \hline\hline
Hill and Reddi~\cite{hill-reddi} &2 -- 3 &camera applications \\ \hline
Bleier et al.~\cite{earable} &1 -- 3 &wearable applications \\ \hline
Bleier et al~\cite{olfactory} &1 -- 3 &olfactory computing \\
 \hline
Karageorgos et al.~\cite{halo} &1 -- 3 &brain-computer interface \\ \hline
\end{tabular}
}
\end{table}

\vspace{1mm}
\noindent{\bf Critical DSA count.}
We now rework the above inequality to:
\begin{equation}
N > n \cdot \left( \frac{E}{A} + \frac{1 - E}{\alpha_{E2O} \cdot A} \right) = \textit{CDC},
\end{equation}
while defining the right-hand side of the inequality as the {\it critical DSA count (CDC)}. This leads to the overall conclusion that if the number of DSAs $N$ is larger than CDC, the reconfigurable fabric incurs a smaller environmental footprint than the sea of DSAs. Because of the conservative assumptions made in terms of DSA/kernel concurrency, we believe that CDC computed using the above formula is an overestimation.

\subsection{Analysis and Discussion}

Despite the simplicity of the model, several interesting conclusions can be obtained by analyzing sensitivity of {\it CDC} with respect to $A$, $E$ and $\alpha_{E2O}$, see Figure~\ref{fig:CDC} assuming (a) serial DSA execution and (b) concurrent DSA execution. $A$ and $E$ are in the range of empirically observed values, see also Section~\ref{sec:results}, with $A = E = 0.35$ close to the average. $n=3$ is typical for DSA concurrency, see Table~\ref{tbl:kernels}. 

First, CDC is a monotonically decreasing curve as a function of $\alpha_{E2O}$, and converges to $n/A$ as the embodied footprint dominates, i.e., $\alpha_{E2O} \rightarrow 1$. This implies that {\em a reconfigurable fabric is especially environmentally friendly for computing devices that are dominated by the embodied footprint, such as battery-operated devices (watches, smartphones, laptops).} Inversely, a sea of DSAs is more environmentally friendly for devices that are dominated by their operational footprint. Which specific design option (a sea of DSAs versus a reconfigurable fabric) is more environmentally friendly depends on the actual difference in area and energy efficiency and the ratio of the embodied versus operational footprint.

Second, area efficiency $A$ has a more significant impact on CDC than energy efficiency $E$. Indeed, the curves are primarily grouped by different values for $A$ and not $E$. This implies --- contrary to common belief perhaps --- that {\em optimizing the area efficiency of a DSA is more important than optimizing its energy efficiency to reduce its overall environmental footprint.} 

Third, the larger the relative area $A$ (and energy consumption $E$) of the DSA compared to the reconfigurable fabric, the smaller the CDC. In other words, {\em if the area and energy reduction of the DSA is relatively small compared to the reconfigurable fabric, the latter leads to a smaller environmental footprint.} This suggests that a reconfigurable fabric is particularly beneficial for sustainability if the area gains through a dedicated accelerator are limited. In contrast, if the DSA is substantially more area-efficient, it is potentially more environmentally friendly than a reconfigurable fabric.

Fourth, {\em a reconfigurable fabric tends to be increasingly more environmentally friendly compared to a sea of a DSAs with limited DSA concurrency.} More concretely, for an embodied-dominated device ($\alpha_{E2O} = 0.8$ and $A=E=0.35$), a reconfigurable fabric incurs a smaller environmental footprint than at most a handful DSAs (serial execution) and a dozen DSAs (concurrent execution). For an operational-dominated device ($\alpha_{E2O} = 0.25$ and $A=E=0.35$), a reconfigurable fabric needs to replace 8 DSAs (serial execution) to 25 DSAs (concurrent execution) for it to be more sustainable. With current SoCs featuring more than 40  DSAs~\cite{hill-reddi}, there appears to be compelling opportunity to improve processor sustainability for replacing the sea of DSAs with a reconfigurable fabric.




\section{Methodology}

We now quantify the typical area and energy efficiency of a DSA relative to a reconfigurable fabric, i.e., $A$ and $E$, for a set of widely used kernels. We first compare the relative area and energy efficiency of dedicated ASIC-based DSAs (as typically implemented in modern-day SoCs) versus a reconfigurable fabric for realizing accelerators. We choose coarse-grain reconfigurable array (CGRA) as the representative reconfigurable fabric. CGRAs are composed of an array of coarse-grained processing elements (PEs) connected via an on-chip interconnect. Both the PEs and the interconnect can be configured to implement different functions and data paths, enabling the acceleration of diverse workloads on a single hardware fabric through software. Our selection of CGRAs is motivated by the fact that CGRAs occupy the middle ground between the flexibility of FPGA versus the efficiency of ASIC. CGRAs provide superior area and energy-efficiency compared to FPGAs~\cite{plasticine} as their basic building blocks for reconfiguration are coarse-grained functional units compared to the individual logic gates of the FPGAs. While not as efficient as the ASICs, CGRAs are flexible to provide excellent acceleration for a wide range of computational kernels.

\subsection{Workloads}
To analyze the carbon footprint between specialization and generalization, we analyze the DSAs and reconfigurable fabrics using a set of workloads from the MachSuite benchmark suite~\cite{machsuite}, see Table~\ref{tbl:benchmarks}. These workloads encompass a wide range of behaviors of frequently accelerated kernels, including applications from linear algebra in machine learning (\textit{GeMM, KNN, Conv2D}), image processing (\textit{Stencil3D}), speech recognition (\textit{Viterbi}), signal processing (\textit{FFT, FIR}), and security (\textit{AES Encryption}). These tasks exhibit a range of memory access patterns, control logic requirements, and arithmetic intensity. Additionally, these kernels constitute a significant portion of the EarBench~\cite{earable} benchmark suite, which comprises applications requiring acceleration in hearing aids, earphones, smart glasses, and similar devices that are dominated by embodied footprint. For realistic workload sizes related to these kernels, we use input data dimensions from the Visual Wake Words Challenge~\cite{visualwake}, which is part of the TinyML Perf benchmark suite~\cite{tinymlperf}.

\subsection{CGRA Modeling}
We design a clustered spatio-temporal CGRA architecture in RTL with an 8x8 processing element (PE) grid, as illustrated in Figure~\ref{fig:Cgra} which is modeled after generic contemporary CGRAs~\cite{Amber, softbrain, hycube}.
Each PE includes a dedicated 32-bit ALU, router/crossbar, and configuration memory. We derive the area and power consumption of the CGRA architecture by synthesizing it on a UMC 40\,nm technology node at 100\,MHz frequency. We map the benchmark suite's kernels onto the CGRA using Morpher~\cite{Morpher}, an open-source specialized compiler for CGRAs, to obtain the performance estimates.  To minimize dynamic power consumption, we use clock gating to disable unused PEs based on the mapping configurations of each kernel. The PEs are connected to a 32-bank memory with enough capacity (256\,KB) to store input and intermediate data for the kernel with the highest data memory requirements in the suite, namely \textit{Stencil3D}, see also Table~\ref{tbl:kernels}.

\begin{table}[t]
\centering
\caption{Application kernels used in the evaluation.}
\label{tbl:benchmarks}
\vspace{-2mm}
\renewcommand{\arraystretch}{0.9}
\resizebox{\columnwidth}{!}{
\scriptsize
\begin{tabular}{|l|l|ll|l|}
\hline
\multicolumn{1}{|c|}{\textbf{App. Kernel}}                & \multicolumn{1}{c|}{\textbf{Domain}}                                          & \multicolumn{2}{c|}{\textbf{Description}}                                                                                       & \multicolumn{1}{c|}{\textbf{\begin{tabular}[c]{@{}c@{}}Memory\\ (in KB)\end{tabular}}} \\ \hline\hline
GeMM                                                      & \multirow{3}{*}{\begin{tabular}[c]{@{}l@{}}Machine \\ Learning\end{tabular}}  & \multicolumn{2}{l|}{\begin{tabular}[c]{@{}l@{}}General Matrix Multiplication \\ 32x32 tile size, 96x96 input size\end{tabular}} & 108                                                                                     \\ \cline{1-1} \cline{3-5} 
KNN                                                       &                                                                               & \multicolumn{2}{l|}{\begin{tabular}[c]{@{}l@{}}K-nearest neighbour\\ 16 maximum neighbours\end{tabular}}                        & 22                                                                                     \\ \cline{1-1} \cline{3-5} 
Conv2D                                                    &                                                                               & \multicolumn{2}{l|}{\begin{tabular}[c]{@{}l@{}}2D convolution\\ Filter size of 3x3, input size 96x96\end{tabular}}              & 72                                                                                   \\ \hline
Stencil3D                                                 & \begin{tabular}[c]{@{}l@{}}Image \\ Processing\end{tabular}                   & \multicolumn{2}{l|}{\begin{tabular}[c]{@{}l@{}}3D stencil calculation with \\ data size 16x32x32\end{tabular}}                  & 256                                                                                    \\ \hline
Viterbi                                                   & \begin{tabular}[c]{@{}l@{}}Speech \\ Recognition\end{tabular}                 & \multicolumn{2}{l|}{\begin{tabular}[c]{@{}l@{}}Viterbi algorithm \\ 64 hidden states 32 observations\end{tabular}}              & 52                                                                                     \\ \hline
FFT                                                       & \multirow{2}{*}{\begin{tabular}[c]{@{}l@{}}Signal \\ Processing\end{tabular}} & \multicolumn{2}{l|}{128-pt fast Fourier transform}                                                                              & 1.5                                                                                    \\ \cline{1-1} \cline{3-5} 
FIR                                                       &                                                                               & \multicolumn{2}{l|}{32-tap FIR filter}                                                                                          & 108                                                                                     \\ \hline
\begin{tabular}[c]{@{}l@{}}AES \\ Encryption\end{tabular} & Security                                                                      & \multicolumn{2}{l|}{Rijndael ciphers with 16B block size}                                                                       & 0.5                                                                                    \\ \hline
\end{tabular}
}
\end{table}

\begin{figure}[tb]
\centering
\includegraphics[width=0.8\columnwidth]{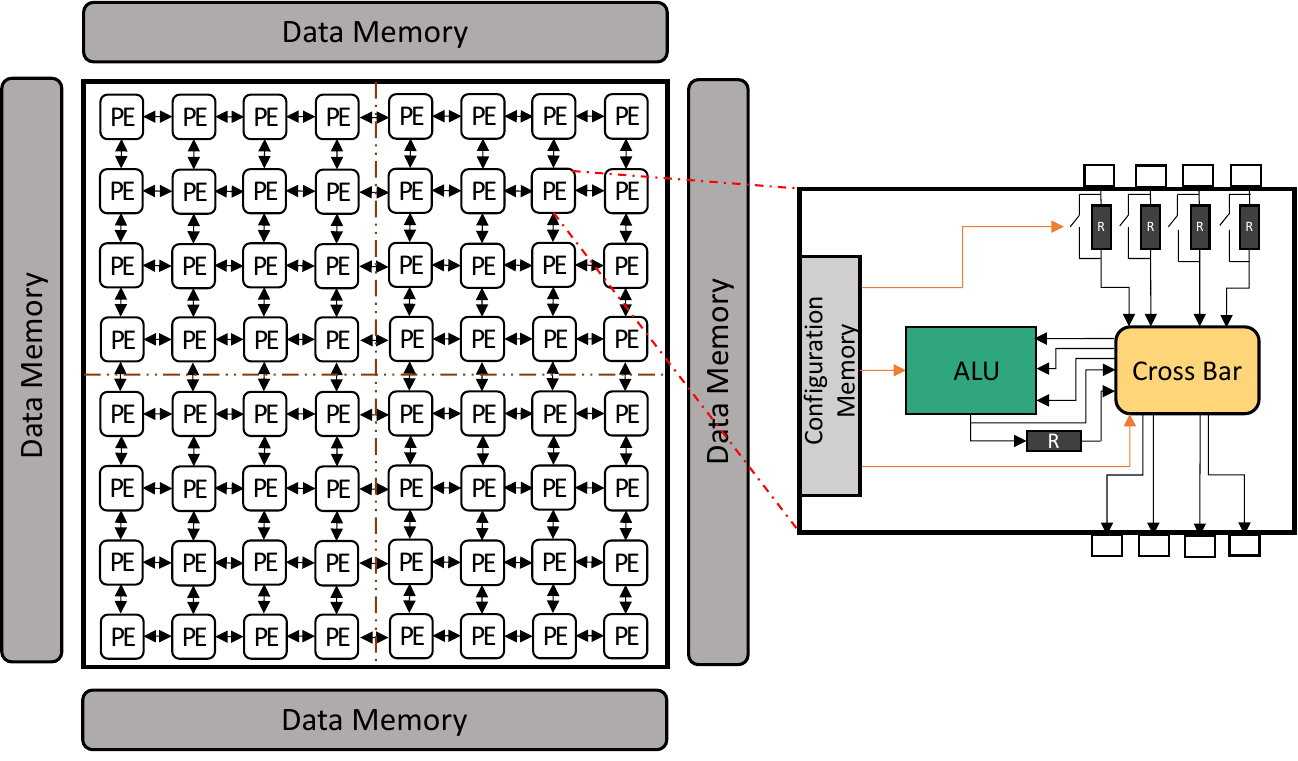}
\vspace{0mm}
\caption{The modeled CGRA architecture.}
\label{fig:Cgra}
\end{figure}

\subsection{DSA Modeling}

To ensure a fair comparison, we explore iso-performance design points for the ASIC-based DSAs for each kernel against the CGRA. It is computationally expensive to identity the iso-performance ASIC-based design point at RTL level. We use Aladdin~\cite{Aladdin}, a modeling tool that allows us to quickly explore a variety of ASIC design points for each kernel, showcasing a range of power, performance and area characteristics without the need for extensive RTL re-writing. 

Aladdin estimates the area and power of the accelerators according to its calibrated data at a 40\,nm technology node, aligning with the CGRA modeling. For consistency, we model the DSAs at 100\,MHz and limit each DSA to 32 SRAM banks, matching the CGRA's maximum on-chip traffic capacity. The on-chip memory requirements of each DSA vary according to the kernel's input and intermediate data storage needs. The DSA for the \textit{Stencil3D} kernel requires the largest data memory, equivalent to that of the CGRA.

We leverage Aladdin to pinpoint design points that closely mirror CGRA performance while prioritizing area and power optimization. This results in DSAs that deviate from their corresponding CGRA kernel implementations by an average of 10\% in performance. 

The selection of our design points is premised on the CGRA (and consequently, the DSAs) adequately meeting the application's performance requirements before optimizing area and power characteristics. Increasing performance further will require higher hardware resources (area, power) in the accelerator architectures. However, it is crucial to assess whether the increase in resource usage translates to a uniform performance gain for both the DSAs and the CGRA.

\begin{figure}[t]
\centering
    \includegraphics[width=0.8\linewidth]{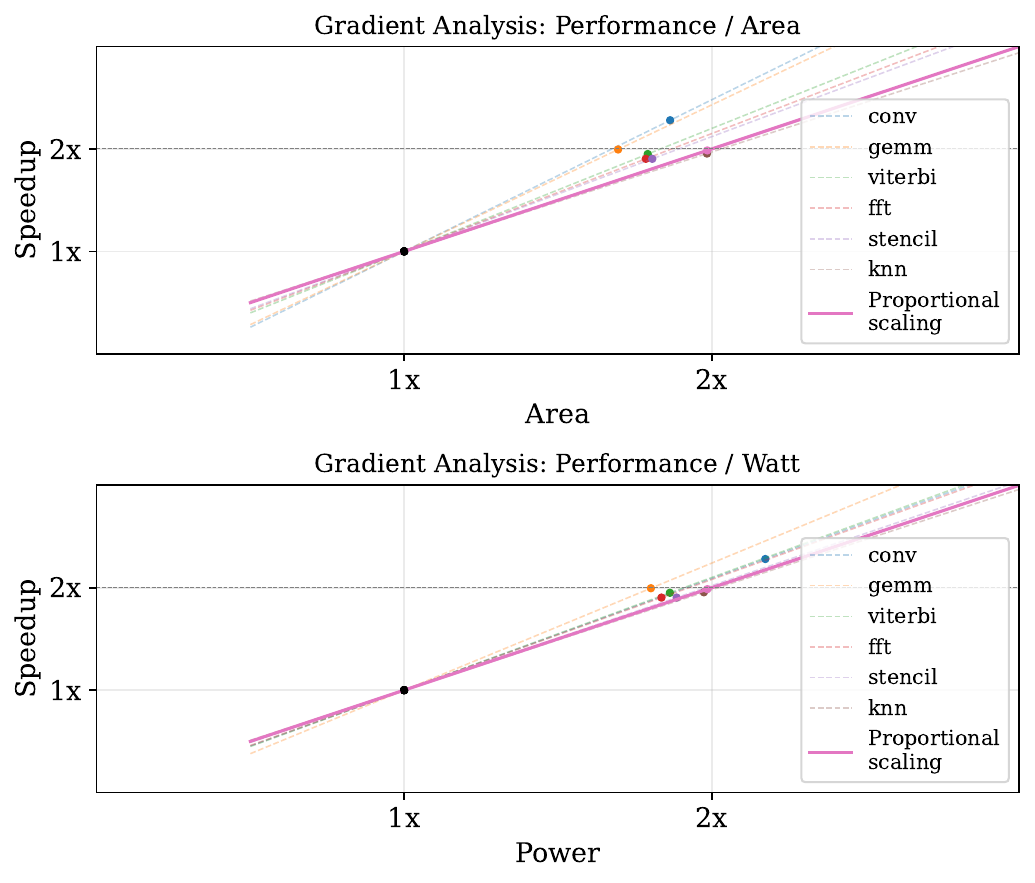}
    \vspace{-5pt}
    \caption{Relationship between performance and resource scaling. {\it The proportional scaling slope indicates performance versus resources for CGRA. The rest of the lines represent performance versus resource scaling for DSAs.}}
    \label{fig: perfarea_slope}
\end{figure}

To evaluate the performance scaling of the DSAs and CGRA with additional area and power, we systematically model the architectures with doubled hardware resources. First, we double the available on-chip memory bandwidth allocated to both the architectures. Then, in the case of CGRA, we double the number of PEs. The CGRA compiler maps to each PE cluster individually, resulting in a 2$\times$ performance improvement as the number of clusters doubles. This proportional scaling is shown in Figure~\ref{fig: perfarea_slope}. 


For each kernel, similar to the baseline 1$\times$ DSA performance point (described previously and used throughout this text), we now find a new, roughly 2$\times$ DSA performance point while optimizing area and power using Aladdin. Subsequently, we plot the new point with the corresponding resource utilizations. We could not find DSA design points for \textit{AES} and \textit{FIR} that are close to 2$\times$ performance.

The robustness of our area and power comparison between DSAs and CGRA relies on a linear relationship between performance and resource scaling. In reality, we observe a steeper increase in performance versus resources for the DSA kernels (slope > 1). Specifically, the median performance versus area slope stands at 1.2, and the median performance versus power slope is 1.09. Aladdin's estimation methodology incorporates optimizations tailored to the specific input data and kernel characteristics, which may contribute to the more favorable performance scaling observed for the DSAs. This contrasts with earlier studies on simulators and real-world accelerators~\cite{stonne, eyeriss}, which show that performance gains do not always match up with the increase in on-chip resources linearly.
Nonetheless, the observed deviations from the proportional scaling line are within a reasonable range. 

Overall, performance scales almost proportionally with hardware resources (i.e., chip area and power) for both DSA and CGRAs. The near-proportional scaling observed across various performance levels suggests that our analysis and conclusions are robust and applicable across a range of performance targets, rather than being limited to a specific operating point. Consequently, the specific iso-performance data point considered in this work does not impact the generality of our findings. We could have chosen a different iso-performance data point for our modeling, and the conclusions would remain largely similar.

\section{Evaluation}\label{sec:results}

The abstract analytical model to compare the environmental footprint of a sea of DSAs versus the CGRA depends on the area $A$ and energy efficiency $E$ of the DSAs relative to the CGRA. As area and energy here serve as proxies for the embodied and operational footprint, it is important to study their impact at an individual level before considering their impact on sustainability.

\begin{figure}[t]
    \centering
    \resizebox{0.9\linewidth}{!}{\begin{tikzpicture}

\begin{axis}  
[  
    ybar = 0pt,
    ylabel near ticks,
    ylabel={Normalized Area},
    label style={font={\fontsize{3.5 pt}{3.5 pt}\selectfont}},
    ticklabel style={font={\fontsize{3.5 pt}{3.5 pt}\selectfont}},
    symbolic x coords={GeMM, FFT, Conv\\2D, Stencil\\3D, FIR, Viterbi, KNN, AES\\Encrypt},
    xtick = data,
    bar width=8pt,
    y post scale = 0.15,
    x post scale = 0.5,
    enlarge x limits = 0.12,
    ymax = 0.6,
    ymin = 0,
    xticklabel style={align=center, rotate=0},
    tick pos = left,
    major tick length=2pt,
]  
\addplot [draw=black!70!white, fill=black!30!white] coordinates {(GeMM, 0.41) (FFT, 0.291) (Conv\\2D, 0.202) (Stencil\\3D, 0.502) (Viterbi, 0.128) (FIR, 0.396) (AES\\Encrypt, 0.03) (KNN, 0.241)};
\end{axis}  
\end{tikzpicture}}
    \vspace{-3mm}
        \caption{Chip area of dedicated DSAs normalized to 8x8 CGRA. {\it A DSA incurs on average $0.27\times$ less area compared to a CGRA with fixed area (not all kernels utilize full CGRA resources.)}}
    \label{fig: Area Comparison}
\end{figure}
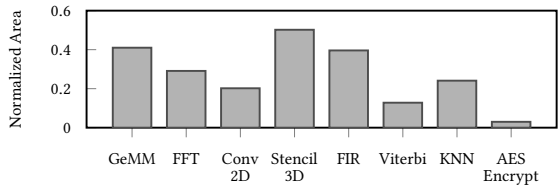 
\vspace*{-3pt}
\subsection{Chip Area}
There has been a lot of effort in the hardware design community on improving device energy efficiencies. As aforementioned, chip area, serving as a proxy for the embodied footprint, turns out to be the more significant metric affecting carbon emissions for computing devices dominated by their embodied footprint (i.e., $\alpha_{E2O} \geq 0.5$). 
Figure~\ref{fig: Area Comparison} reports chip area for the various DSAs relative to the CGRA. The DSAs consume between $0.03\times$ and $0.55\times$ compared to the fixed 8x8 CGRA area, with an average of $0.27\times$ across all kernels. 

\begin{figure}[t]
    \centering
    \resizebox{0.9\linewidth}{!}{\begin{tikzpicture}

\begin{axis}  
[  
    ybar = 0pt,
    ylabel near ticks,
    ylabel={Normalized Energy},
    label style={font={\fontsize{3.5 pt}{3.5 pt}\selectfont}},
    ticklabel style={font={\fontsize{3.5 pt}{3.5 pt}\selectfont}},
    symbolic x coords={GeMM, FFT, Conv\\2D, Stencil\\3D, FIR, Viterbi, KNN, AES\\Encrypt},
    xtick = data,
    bar width=8pt,
    y post scale = 0.15,
    x post scale = 0.5,
    enlarge x limits = 0.12,
    ymax = 0.6,
    ymin = 0,
    xticklabel style={align=center, rotate=0},
    tick pos = left,
    major tick length=2pt,
]  
\addplot [ybar, draw=black!70!white, fill=black!30!white] coordinates {(GeMM, 0.541) (FFT, 0.283) (Conv\\2D, 0.410) (Stencil\\3D, 0.511) (Viterbi, 0.091) (FIR, 0.395) (AES\\Encrypt, 0.04) (KNN, 0.479)};
  
\end{axis}  
\end{tikzpicture}}
    \vspace{-4mm}
        \caption{Energy consumption of dedicated per-kernel DSAs normalized to the CGRA. {\it A DSA consumes on average $0.31\times$ less energy compared to a CGRA.}}
    \label{fig: Energy Comparison}
\end{figure}
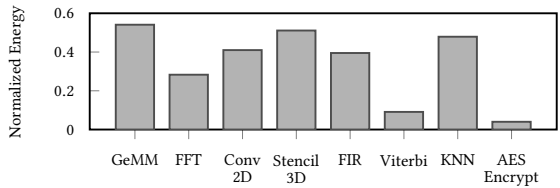

CGRAs feature a fixed number of PEs and, consequently, a fixed number of Arithmetic Logic Units (ALUs) for computations. ALUs are versatile and capable of executing various operations such as multiplication, accumulation, shift, and bitwise operations. It is important to note that the area required for a multiplication operation significantly exceeds that of a bitwise operation. Further, CGRAs have complex, resource-expensive crossbars and routers to handle various kinds of dataflows. This observation aligns with the findings in Figure~\ref{fig: Area Comparison} and Figure~\ref{fig: Energy Comparison}, where DSAs involving a lot of multiply-accumulate operations, like \textit{GeMM, Stencil3D} and \textit{FIR}, exhibit significantly higher area and energy utilization compared to others. 
In turn, this also results in noticeably lower resource consumption for DSAs like \textit{AESEncrypt} and \textit{Viterbi}, which primarily rely on bitwise operations, lookups, or iterative updates. 

\subsection{Energy Consumption}
The total energy consumption of the architecture for the duration of its operation serves as a proxy for the operational footprint in the model. Our analysis in Figure~\ref{fig: Energy Comparison} indicates that the DSAs consume between $0.03 \times$ to $0.49\times$ less energy compared to the CGRA, with an average of $0.31\times$ across all kernels. 

For compute-intensive kernels with a high density of multiply-accumulate operations, such as \textit{GeMM}, \textit{Stencil3D} and \textit{FIR}, DSAs consume slightly more than half the energy compared to the corresponding CGRA implementation of those kernels. Conversely, for tasks with irregular loops like {\em Viterbi} and complex routing like {\em FFT}, DSAs demonstrate notably lower energy consumption. This advantage stems from the challenges CGRAs face in efficiently routing and mapping these types of applications onto their architecture. Furthermore, the CGRA's inherent flexibility becomes overkill for such tasks with moderate computational demands. The reconfigurability and large ALUs of the CGRA result in significantly higher energy usage compared to DSAs specifically designed for these applications without the extra features.

\subsection{Critical DSA Count}\label{sec:CDC}

We now explore the potential for replacing DSAs with a CGRA from a sustainability perspective. If the number of DSAs (i.e., $N$) surpasses the Critical DSA Count (\textit{CDC}), then the CGRA becomes the preferable and more sustainable option. 


The CDC demonstrates considerable variation influenced by the area, performance, and power characteristics of the DSAs anticipated to be substituted by a CGRA. Considering this, we undertake a comprehensive analysis of various scenarios as case studies, aimed at evaluating the relative impact of DSA types and their characteristics. Through these case studies, we seek to provide insight into the decision-making process and feasibility considerations associated with transitioning from a sea of accelerators to a CGRA. We consider the following scenarios:
\begin{itemize}
    \item \textbf{CASE-I:} \textbf{All DSAs.} In this scenario, we examine the CDC when all DSAs are potentially up for replacement by a CGRA. This scenario works without detailed insights into the specific characteristics of individual DSAs.
    \item \textbf{CASE-II:} \textbf{All DSAs minus AES.} Here we analyze the CDC in a scenario where we exclude only the smallest outlier kernel (\textit{AESEncrypt}) from consideration for replacement. This approach proposes a case for leaving the smallest outlier kernel as an individual DSA on the chip (because it consumes miniscule area) while replacing all other kernels. 
    \item \textbf{CASE-III:} \textbf{All DSAs minus AES/Viterbi.} Finally, we extend CASE-II to exclude the two small outlier kernels (\textit{AESEncrypt} and \textit{Viterbi}) from consideration for replacement. This approach prioritizes the replacement of larger kernels with a CGRA while retaining all smaller kernels on the chip as individual DSA accelerators with minimal impact on area.
\end{itemize}

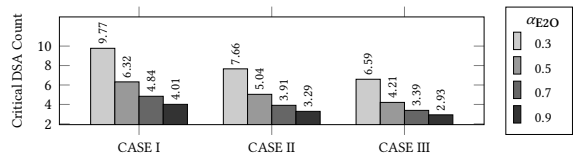
\begin{figure}[t]
    \centering
    \resizebox{0.9\linewidth}{!}{\begin{tikzpicture}  
\centering
\begin{axis}  
[  
    ybar, 
    ylabel near ticks,
    nodes near coords,
    every node near coord/.append style={font={\fontsize{5 pt}{5 pt}\selectfont}},
    every node near coord/.append style={rotate=90, anchor=west, xshift=-0.05cm},
    ylabel={Critical DSA Count},
    label style={font={\fontsize{6 pt}{6 pt}\selectfont}},
    symbolic x coords={CASE I, CASE II, CASE III}, 
    ticklabel style={font={\fontsize{6 pt}{6 pt}\selectfont}},
    xtick=data,
    legend pos=north east,
    bar width=10pt,
    ybar = 0pt,
    ymax=13,
    y post scale = 0.275,
    x post scale = 0.9,
    enlarge x limits=0.3,
    ytick = {2, 4, 6, 8, 10},
    legend style={font={\fontsize{6 pt}{6 pt}\selectfont}},
    legend style={/tikz/every even column/.append style={column sep=0.2cm}},
    legend image post style={xscale=1, yscale=1},
    legend columns = 1,
    legend style={at={(1.05,0.5)},anchor=west},
    legend image code/.code={
                \draw [/tikz/.cd,bar width=3pt,yshift=-0.2em,bar shift=0pt]
                plot coordinates {(0cm,0.8em)};
            },
    xticklabel style={align=left, rotate=0},
]  
\addlegendimage{empty legend}

\addplot [draw=black,fill=black!20!white] coordinates {(CASE I, 9.773) (CASE II, 7.66) (CASE III, 6.59)};

\addplot [draw=black,fill=black!40!white] coordinates {(CASE I, 6.32) (CASE II, 5.04) (CASE III, 4.21)};

\addplot [draw=black,fill=black!60!white] coordinates {(CASE I, 4.84) (CASE II, 3.91) (CASE III, 3.39)};

\addplot [draw=black,fill=black!80!white] coordinates {(CASE I, 4.01) (CASE II, 3.29) (CASE III, 2.93)};

\legend{$\mathbf{\alpha_{E2O}}$, 0.3, 0.5, 0.7, 0.9}  

\end{axis}  
\end{tikzpicture}  }
    \vspace{-4mm}
    \caption{CDC for scenarios I, II and II for different values of $\alpha_{E2O}$; no concurrency ($n=1$). {\it CDC decreases with increasing $\alpha_{E2O}$ (higher contribution of embodied footprint) and decreasing area efficiency of the DSAs relative to the CGRA (increasing $A$ corresponding to scenarios I through III).}}
    \label{fig: CDC Evaluations}
\end{figure}

Figure~\ref{fig: CDC Evaluations} reports CDC for the three cases while considering different values of $\alpha_{E2O}$. Our analysis focuses mainly on $\alpha_{E2O} \geq 0.5$ which is the region where embodied carbon starts dominating over operational. Battery-operated devices typically fall in the region of $ 0.7 \leq \alpha_{E2O} \leq 0.9$. We assume serial DSA execution for now; we consider DSA concurrency in the next subsection.

\vspace{1mm} {\bf CASE-I:}
Figure~\ref{fig: CDC Evaluations} demonstrates that even under the most conservative conditions for battery-operated devices ($\alpha_{E2O} = 0.7$),  merely 4--5 accelerators already account for a carbon footprint equivalent to that of a CGRA. As this case overlooks the specifics of individual DSA characteristics, it serves as a valuable approach for a blanket replacement of all DSAs with a CGRA on a chip. Furthermore, its applicability extends across a broader spectrum of scenarios for sustainability-focused architectural decisions.


\vspace{1mm}{\bf CASE-II:} 
As depicted in Figure~\ref{fig: Area Comparison}, the dedicated DSA for \textit{AESEncrypt} kernel occupies an extremely small footprint compared to the other DSAs. Simply excluding this kernel from consideration for replacement leads to a substantial reduction in CDC. Figure~\ref{fig: CDC Evaluations} reports that the CDC for CASE-II in the region dominated by embodied emissions falls below 4. Just the exclusion of one small kernel makes a noticeable difference in the replaceability of DSAs with a reconfigurable fabric, emphasizing the need for insight into individual DSA characteristics before swapping them out for a CGRA.

\vspace{1mm}{\bf CASE-III:}
The analysis in Case II underscores the importance of further studying the impact of excluding smaller DSAs from the replacement pool. DSAs corresponding to the \textit{Viterbi} and \textit{AES Encrypt} kernels are the small outlier kernels within the benchmark suite, as shown by the earlier energy and area comparisons. By targeting only the larger-footprint kernels for replacement, we observe a further reduction in CDC. In fact, for $\alpha_{E2O} = 0.9$, the CDC dips below 3. This suggests that sustainability benefits could be realized by replacing as few as three DSAs with a CGRA architecture. The analysis emphasizes the importance of selecting the appropriate set of dedicated DSAs for replacement with a reconfigurable fabric to minimize the carbon footprint.


\subsection{CDC for Concurrent Execution}\label{scaling}





\begin{figure*}[t]
\subfloat[\small $n$ = 2]{
\centering
\begin{tikzpicture}
\begin{axis}[
    height=5cm,width=0.35\textwidth,
    bar width=7pt,
    ybar, 
    ylabel near ticks,
    nodes near coords,
    every node near coord/.append style={font={\fontsize{5 pt}{5 pt}\selectfont}},
    every node near coord/.append style={rotate=90, anchor=west, xshift=-0.05cm},
    ylabel={Critical DSA Count},
    label style={font={\fontsize{6 pt}{6 pt}\selectfont}},
    symbolic x coords={CASE-I, CASE-II, CASE-III}, 
    ticklabel style={font={\fontsize{6 pt}{6 pt}\selectfont}},
    xtick=data,
    ymax=25,
    legend pos=north east,
    ybar = 0pt,
    y post scale = 0.615,
    x post scale = 0.9,
    enlarge x limits=0.3,
    xticklabel style={align=left, rotate=0},
]
\addlegendimage{empty legend}
\addplot [fill=black!20!white] table [x=Case, y=0.3, col sep=space] {data_2.csv};
\addplot [fill=black!40!white] table [x=Case, y=0.5, col sep=space] {data_2.csv};
\addplot [fill=black!60!white] table [x=Case, y=0.7, col sep=space] {data_2.csv};
\addplot [fill=black!80!white] table [x=Case, y=0.9, col sep=space] {data_2.csv};
\end{axis}
\end{tikzpicture}
}
\hspace{1mm}
\subfloat[\small $n$ = 3]{
\centering
\begin{tikzpicture}
\begin{axis}[
    height=5cm,width=0.35\textwidth,
    bar width=7pt,
    ybar, 
    ylabel near ticks,
    nodes near coords,
    every node near coord/.append style={font={\fontsize{5 pt}{5 pt}\selectfont}},
    every node near coord/.append style={rotate=90, anchor=west, xshift=-0.05cm},
    ylabel={Critical DSA Count},
    label style={font={\fontsize{6 pt}{6 pt}\selectfont}},
    symbolic x coords={CASE-I, CASE-II, CASE-III}, 
    ticklabel style={font={\fontsize{6 pt}{6 pt}\selectfont}},
    xtick=data,
    ymax=25,
    legend pos=north east,
    ybar = 0pt,
    y post scale = 0.615,
    x post scale = 0.9,
    enlarge x limits=0.3,
    xticklabel style={align=left, rotate=0},
]
\addlegendimage{empty legend}
\addplot [fill=black!20!white] table [x=Case, y=0.3, col sep=space] {data_3.csv};
\addplot [fill=black!40!white] table [x=Case, y=0.5, col sep=space] {data_3.csv};
\addplot [fill=black!60!white] table [x=Case, y=0.7, col sep=space] {data_3.csv};
\addplot [fill=black!80!white] table [x=Case, y=0.9, col sep=space] {data_3.csv};
\end{axis}
\end{tikzpicture}
}
\hspace{1mm}
\subfloat[\small $n$ = 4]{
\centering
\begin{tikzpicture}
\begin{axis}[
    height=5cm,width=0.35\textwidth,
    bar width=7pt,
    ybar, 
    ylabel near ticks,
    nodes near coords,
    every node near coord/.append style={font={\fontsize{5 pt}{5 pt}\selectfont}},
    every node near coord/.append style={rotate=90, anchor=west, xshift=-0.05cm},
    ylabel={Critical DSA Count},
    label style={font={\fontsize{6 pt}{6 pt}\selectfont}},
    symbolic x coords={CASE-I, CASE-II, CASE-III}, 
    ticklabel style={font={\fontsize{6 pt}{6 pt}\selectfont}},
    xtick=data,
    legend pos=north east,
    ybar = 0pt,
    y post scale = 0.615,
    x post scale = 0.9,
    enlarge x limits=0.3,
    ymax=25,
    legend style={font={\fontsize{6 pt}{6 pt}\selectfont}},
    legend style={/tikz/every even column/.append style={column sep=0.2cm}},
    legend image post style={xscale=1, yscale=1},
    legend columns = 1,
    legend style={at={(1.15,0.5)},anchor=west, rotate=90},
    legend image code/.code={
                \draw [/tikz/.cd,bar width=3pt,yshift=-0.2em,bar shift=0pt]
                plot coordinates {(0cm,0.8em)};
            },
    xticklabel style={align=left, rotate=0},
]
\addlegendimage{empty legend}
\addplot [fill=black!20!white] table [x=Case, y=0.3, col sep=space] {data_4.csv};
\addplot [fill=black!40!white] table [x=Case, y=0.5, col sep=space] {data_4.csv};
\addplot [fill=black!60!white] table [x=Case, y=0.7, col sep=space] {data_4.csv};
\addplot [fill=black!80!white] table [x=Case, y=0.9, col sep=space] {data_4.csv};
\legend{$\mathbf{\alpha_{E2O}}$, 0.3, 0.5, 0.7, 0.9}
\end{axis}
\end{tikzpicture} 
}
\vspace{-2mm}
\caption{Sensitivity of CDC with respect of DSA characteristics (CASE I, II, III), $\mathbf{\alpha_{E2O}}$ (legend) and kernel concurrency (subfigures (a), (b) and (c)). {\it CDC decreases with (1) reduced area and energy efficiency of the DSA (from CASE-I to CASE-III), (2) increased weight of the embodied footprint ($\alpha_{E2O}$) in the total footprint, and (3) decreasing kernel concurrency ($n$).}}
\label{fig: CDC Concurrency Evaluations}
\end{figure*}
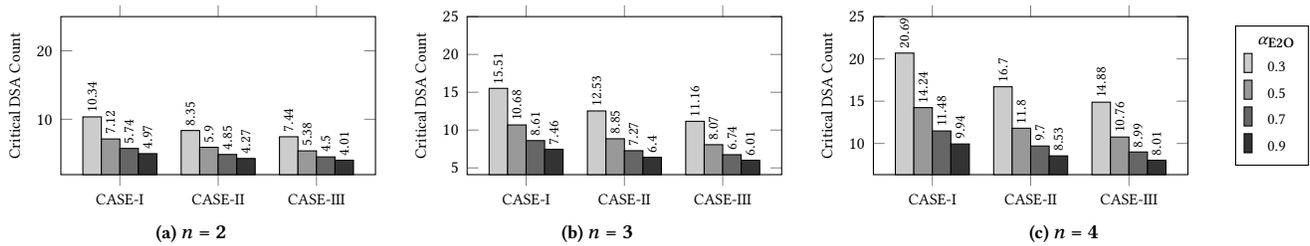

For the above analysis, we focused on applications with only one active DSA ($n=1$), which suits applications requiring sequential execution of various kernels. However, when $n>1$, i.e., applications where multiple DSAs are required to be active simultaneously, the area or number of CGRAs must increase to accommodate multiple kernels while maintaining consistent performance. The worst-case scenario occurs when each kernel utilizes 100\% of the CGRA resources, necessitating a proportional area increase by a factor $n$. However, most kernels do not fully utilize the computational resources of the CGRA. Out of the kernels considered in this study, \textit{GeMM} and \textit{FIR} utilize 100\% of the compute resources, while resource utilization is less than 50\% for \textit{Conv2D}, \textit{Stencil3D}, \textit{Viterbi}, and \textit{AES Encryption}, resulting in an average utilization of $64\%$. 

This allows multiple kernels to utilize the CGRA fabric simultaneously. To analyze the sustainability advantages for concurrent kernels, we scale up the CGRA footprint according to this average utilization to accommodate $n$ kernels in parallel. This scaling up enables low resource usage kernels to compensate for higher resource kernels when running in parallel on CGRA. For Case I and $n = \{2,3,4\}$, the CGRA requires a scaling up factor of $n' = \{ 1.3\times, 1.9\times,  2.6\times \}$ for the CGRA and DSAs to provide identical performance. $n'$ values are slightly higher for Case II and Case III.

Figure~\ref{fig: CDC Concurrency Evaluations} demonstrates CDC across various levels of concurrency for the three scenarios outlined in Section~\ref{sec:CDC}. Each graph compares CDC for four different values of $\alpha_{E2O}$, with $\alpha_{E2O}=0.3$ representing operational-footprint dominance and $\alpha_{E2O}=0.9$ representing embodied-footprint dominance. The number of active DSAs increases in correlation to concurrency, increasing CDC for all $\alpha_{E2O}$ values. In the case of $\alpha_{E2O}=0.7$, where embodied carbon emissions start to dominate, the reconfigurable fabric would be equivalent to 9 to 12 DSAs in carbon emissions for these cases. 

Furthermore, the reconfigurable fabric enhances data locality by mapping the data-dependent accelerators closer and sharing memory storage  while DSAs necessitate frequent data transfers between accelerators. This can lead to a further increase in the energy consumption of the DSAs executing concurrent kernels, which is not quantified in this result. 

\subsection{Environmental Footprint Savings}
While the Critical DSA Count provides valuable insights into the decision-making process for replacing a sea of DSAs with a CGRA, we aim to quantify the carbon footprint reduction associated with replacing a sea of (say tens of) DSAs on a chip with multiple CGRAs, while considering concurrency and performance requirements.

Modern SoCs have upwards of 40 DSAs on a single chip~\cite{aladdin-retrospective}, however, strict thermal constraints often prevent the simultaneous use of all these accelerators on-chip, resulting in dark silicon~\cite{dark-silicon}. By combining this understanding with knowledge about the anticipated concurrency levels of kernels in applications, we can strategically replace the array of tens of accelerators with a smaller set of CGRAs. This approach not only optimizes resource utilization but also enhances sustainability by mitigating dark silicon and reducing the carbon footprint. Our analysis assumes a representative chip with 40 DSAs following prior work and concurrency of up to 5.  


\begin{table}[h]
\centering
\renewcommand{\arraystretch}{0.95}
\caption{Carbon footprint savings by replacing DSAs with CGRA at different concurrency levels.} 
\label{tab:savings}
\vspace{-3mm}
\resizebox{\columnwidth}{!}{
\scriptsize
\begin{tabular}{|c|cc|}
\hline
\multirow{2}{*}{\textbf{\begin{tabular}[c]{@{}c@{}}Concurrency\\ ($n$)\end{tabular}}} & \multicolumn{2}{c|}{\textbf{Carbon Footprint Improvement}}                                        \\ \cline{2-3} 
                                                                                  & \multicolumn{1}{c|}{\textbf{Avg Util ($n' \textless  n$)}} & \textbf{100\% Util ($n' = n$)} \\ \hline\hline
1                                                                                 & \multicolumn{1}{c|}{- }                                         & $7.60\times$                        \\ \hline
2                                                                                 & \multicolumn{1}{c|}{$6.10\times$}                                          & $3.84\times$                         \\ \hline
3                                                                                 & \multicolumn{1}{c|}{$4.12\times$}                                          & $2.59\times$                         \\ \hline
4                                                                                 & \multicolumn{1}{c|}{$3.12\times$}                                          & $1.97\times$                         \\ \hline
5                                                                                 & \multicolumn{1}{c|}{$2.53\times$}                                          & $1.59\times$                         \\ \hline
\end{tabular}
}
\end{table}

\vspace{-2pt}


Table~\ref{tab:savings} quantifies the improvement in carbon footprint achieved by using CGRAs instead of a sea of DSAs. CGRAs achieve between 2.53 $\times$ and 7.6 $\times$ less carbon footprint when considering the average utilization of the hardware resources. Here, $n$ denotes the expected concurrency level in the application domain for the chip, while $n'$ represents the scaling-up factor for a CGRA to accommodate such concurrency in applications with average resource utilization (see Section \ref{scaling}). For a chip where only sequential execution is anticipated in applications, replacing all DSAs with CGRAs leads to a substantial 7.6$\times$ improvement in carbon footprint. This saving over a sea of accelerators decreases as we start working with applications requiring higher levels of concurrency. As previously discussed, many applications exhibit concurrency levels ranging from 1 to 3, or even up to 4. Even under such expected levels of concurrency, the choice remains clear: Even in the worst-case scenario, utilizing CGRAs results in close to a 2$\times$ reduction in carbon footprint, turning out to be significantly better from the point of view of sustainability. The benefit is slightly lower when every kernel demonstrates 100\% utilization of the CGRA, an unlikely scenario.


Building on the CDC replacement discussion, frequently used small DSAs (e.g., {\em AESEncrypt}) present a unique opportunity. Given its small footprint and prevalence across applications, maintaining {\em AESEncrypt} as a dedicated DSA while substituting other DSAs with CGRAs could further reduce the carbon footprint, especially at higher concurrency, such as $n = 4$. At this concurrency level, keeping \textit{AESEncrpyt} as a separate DSA alongside CGRAs, compared to a chip with 40 DSAs, results in a $4.05\times$ better carbon footprint in contrast to the $3.12\times$ improvement for $n=4$ observed in Table~\ref{tab:savings} previously.  This is because this approach allows for a smaller CGRA for the remaining 3 concurrent kernels, alongside the compact DSA, leading to a more efficient overall design (Case II in Section \ref{sec:CDC}). 
\subsection{Second-Order Benefits}
While we have extensively discussed the first-order sustainability benefits of using reconfigurable fabrics for hardware specialization, there exist some second-order benefits that are difficult to quantify. When multiple DSAs execute within an application, significant data movement occurs across DSAs and their memory hierarchies. CGRAs can multiplex kernels in space and time, reducing such data movement. Moreover, reconfigurable architectures can be manufactured on a larger scale, agnostic of the application/device/company it will serve, streamlining manufacturing by producing larger quantities of identical chips instead of smaller quantities of different ones. Additionally, designing highly specific accelerators for an application severely limits the chip's adaptability to changes. However, with reconfigurable fabrics, new algorithms, and kernels can be seamlessly integrated, optimizations can be made to existing kernels, and entirely new applications can be supported through simple `software updates'. This flexibility prolongs the device's lifespan and usability, thereby enabling sustainable hardware specialization.


\ignore{

\section{Discussion}

\remark{I think it makes sense to have a discussion section that talks about other benefits for reconfigurable logic not exposed in the analysis, such as the possibility to add new kernels on the fly (possibly extending the lifetime of the device because this results in merely a software update and does not require changing hardware). Also, there is somewhere else in the paper a paragraph about the data transfer cost in terms of energy between DSAs, which you eliminate with the CGRA -- this is not accounted for in our analysis, making it a conservative estimate. -- Lieven}

\remark{ I'm not sure about the server-scale specialization. I don't see the point of adding Fig 10 as it is similar to an earlier graph in the paper. So that subsection I would remove. Thanks. -- LIeven}

\subsection{Specialization at server-scale}
\begin{figure}[tb]
\centering
\includegraphics[width=1\columnwidth]{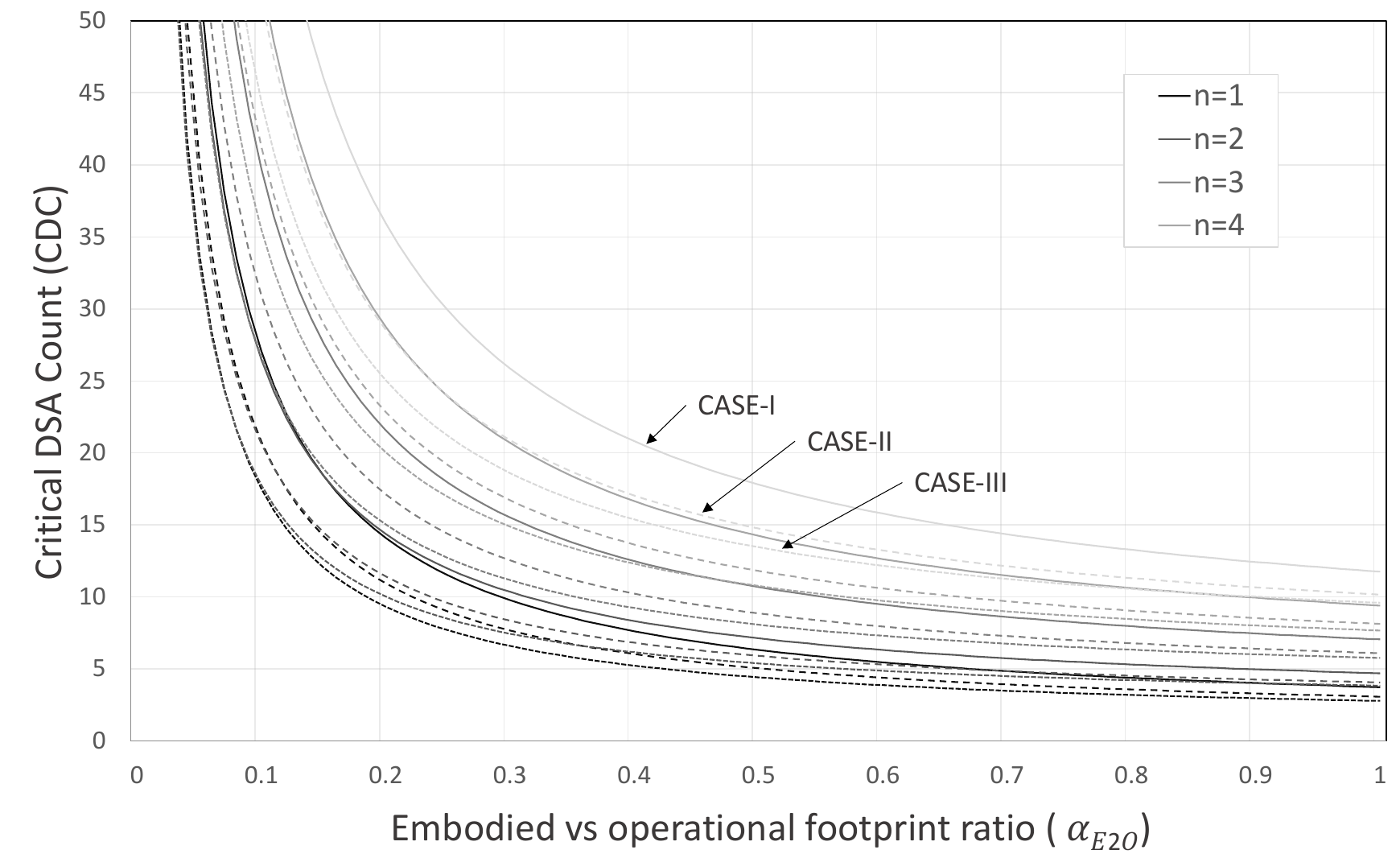}
\vspace{-4mm}
\caption{Variation of CDC with $\alpha_{E2O}$ for different embodied-operational regions}
\label{fig:Cgra}
\vspace{-3mm}
\end{figure}

}

\section{Related Work}
Reconfigurable accelerators have emerged as an alternative to reduce the non-recurring engineering costs associated with ASICs. Ours is the first work to motivate reconfigurable logic as a substitute for a sea of DSAs from a sustainability perspective in the dark silicon era. We focus on CGRA instead of FPGA because of the former's superior area and energy efficiency. 

While FPGAs are the most widely used reconfigurable accelerators, their efficiency is hindered by the higher power and area requirements resulting from their bit-level reconfigurability. According to Kuon and Rose~\cite{fpga_asic_gap}, the area of an FPGA is $40\times$ higher than that of an ASIC in a logic-only FPGA, and it is $21\times$ higher in a modern FPGA with logic, DSPs, and BRAMs. Wong et al. ~\cite{fpga_cmos} show that an FPGA sees a $17\times$ - $27\times$ area increase compared to a custom CMOS implementation. Kuon and Rose~\cite{fpga_asic_gap} also compared the dynamic power dissipation, which is $9\times$ higher in modern FPGAs. This significantly reduces the viability of FPGAs as a sustainable computing alternative, particularly for battery-operated devices where embodied footprint dominates.

CGRA, with a coarser word-level reconfigurability, offers an efficient alternative to FPGAs, bringing its area and power consumption closer to that of ASICs. Numerous CGRAs are proposed in industry~\cite{srp,intel,dark-silicon-harmful,renesas} and academia~\cite{hycube,softbrain,adres,snafu}. SoftBrain~\cite{softbrain} CGRA shows that its area and energy are within $8\times$ and $2\times$ of the ASIC values, respectively. Performance \cite{softbrain, cascade, fifer, plasticine} and energy efficiency ~\cite{snafu, riptide, revamp,ultra_elastic,flex} are primary optimization criteria in modern CGRA design. From a sustainability standpoint,  area efficiency of CGRAs is equally crucial, particularly in edge devices where the embodied footprint plays a significant role.


\section{Conclusion}

Dark silicon fundamentally trades off chip area for power efficiency, which is not environmentally sustainable due to its increasing embodied footprint as technology evolves. This work explored a sustainable alternative through reconfigurable logic. Although reconfigurable logic incurs a larger operational footprint, it significantly reduces the embodied footprint by amortizing chip area across many kernels. By combining abstract analytical modeling and hardware synthesis, we find that a representative reconfigurable fabric, namely CGRA, can drastically reduce the environmental footprint of hardware specialization compared to a sea of DSAs.



\begin{acks}
    This research is partially supported by Research Foundation Flanders (FWO) grant no. G018722N and by the National Research Foundation, Singapore, under its Competitive Research Programme Award NRF-CRP23-2019-0003.
\end{acks}

\newpage
\balance
\bibliographystyle{ACM-Reference-Format}
\bibliography{references}

\end{document}